\documentclass[aps,preprint]{revtex4}
\usepackage{graphicx}

\begin{document}

\title{Simplifying Random Satisfiability Problem by Removing Frustrating Interactions}

\author{A. Ramezanpour} \email{ramezanpour@iasbs.ac.ir}

 \affiliation{Institute for Advanced Studies in Basic Sciences,
Zanjan 45195-1159, Iran}

\author{S. Moghimi-Araghi}
\email{samanimi@sharif.edu}

\affiliation{Department of Physics, Sharif University of
Technology, P.O.Box 11365-9161, Tehran, Iran}

\date{\today}

\begin{abstract}
How can we remove some interactions in a constraint satisfaction
problem (CSP) such that it still remains satisfiable? In this paper
we study a modified survey propagation
algorithm that enables us to
address this question for a prototypical CSP, i.e. random
K-satisfiability problem. The average number of removed
interactions is controlled by a tuning  parameter in the
algorithm. If the original problem is satisfiable then we are able to
construct satisfiable subproblems ranging from the original one to
a minimal one with minimum possible number of interactions.
The minimal satisfiable subproblems will provide directly
the solutions of the original problem.
\end{abstract}

 \maketitle

\section{Introduction}\label{1}
There are many combinatorial problems that can be represented as a
constrained satisfaction problem (CSP) in which we are to satisfy
a number of constrains defined over a set of discrete variables.
An interesting example is the Low Density Parity Check Code in
information theory \cite{ga}. Here a code word consists of $N$
variables $\in \{0,1\}$ that satisfy $M$ parity-check constraints.
Each constraint acts on a few variables and is satisfied if sum of
the variables module $2$ is zero. Another example is finding the
fixed points of a Random Boolean Network \cite{clpwz}. Again we
have $N$ Boolean variables represented by the nodes of a directed
network. The state of a node at a given time step is a logical
function of the state of its incoming neighbors in the previous
time step. Thus a fixed point of the problem is one that satisfies
$N$ constraints, one for each variable, where a constraint
enforces the variable taking the outcome of the logical function.\\
From a physical point of view there exist a close relation between
these problems with frustrated systems exhibiting glassy behavior,
such as spin glasses \cite{mpv}. The methods and concepts
developed in the study of these systems enable us to
get a better understanding of the above problems.\\
Random satisfiability problem is a typical CSP that allows us to
study combinatorial CSP's in a simple framework. It is the first
problem whose NP-completeness has been proven \cite{co,pa}.
The problem is defined over $N$ logical variables that
are to satisfy $M$ logical constraints or clauses. Each clause
interacts with some randomly selected variables that can appear
negated or as such with equal probability. The clause is satisfied
if at least one of the variables are TRUE. Here the interest is in
the satisfiability of the problem and finding the solutions or
ground state configurations that result to the minimum number of
violated clauses. For small number of clauses per variable
$\alpha=M/N$, a typical instance of the problem is satisfiable,
that is there is at least one configuration of variables that
satisfies all the clauses. On the other hand, for large $\alpha$ a
typical instance of the problem is unsatisfiable with probability
one. We have a sharp transition at $\alpha_c$ that separates SAT
and UNSAT phases of the problem \cite{ks}.\\
The interaction pattern of clauses with variables make a graph
that is called the factor graph \cite{kfl}. Notice that larger
number of interactions lead to much frustration and thus make
the problem harder both in checking its satisfiability and finding
its solutions. Therefore, one way to make the problem easier is to
reduce it to some smaller subproblems with smaller number of
interactions. Then we could utilize some local search algorithms
(like Walksat and its generalizations \cite{al}) to solve the
smaller subproblem. However, for a given number of
variables and clauses the chance to find a solution decreases as we remove
the interactions from the factor graph. Moreover, the number of
subproblems with a given number of interactions is exponentially large.
These facts make the above reduction procedure inefficient unless
we find a way to get around them.\\
Survey propagation algorithm is a powerful massage passing
algorithm that helps us to check the satisfiability of the problem
and find its solutions \cite{mez,bmz}. In Ref. \cite{rm} we showed
that as long as we are in the SAT phase we can modify this
algorithm to find the satisfiable spanning trees. There, we also
showed that there is a correspondence between the set of solutions
in the original problem and those of the satisfiable spanning trees.
Indeed the modified algorithm enabled us to remove some
interactions
from the problem such that the obtained subproblem is still satisfiable.\\
In this paper we are going to investigate the modified algorithm
in more details, by studding its performance for different classes
of subproblems. There is a free parameter in the algorithm that
allows us to control the number of interactions in the
subproblems. In this way we can construct ensembles of satisfiable
subproblems with different average number of interactions. The
largest subproblem is the original problem and the smallest one is
a subproblem in which each clause interacts with just one
variable. The latter satisfiable subproblems, which we call
minimal satisfiable subproblems, result directly to the solutions
of the original problem. We will show how the number of solutions
(in replica symmetric approximation) and the complexity (in
one-step replica symmetric approximation) varies for different
subproblems close to the SAT-UNSAT transition.\\

The paper is organized in this manner: First we define more
precisely the random K-satisfiability problem and its known
features. In section \ref{3} we briefly introduce belief and
survey propagation algorithms that play an essential role in the
remaining parts of the paper. Section \ref{4} has been divide to
four subsections that deal with satisfiable subproblems. We start
by some general arguments and then represent numerical results for
different satisfiable subproblems. Finally section \ref{5} is
devoted to our conclusion remarks.

\section{Random K-satisfiability problem }\label{2}
A random satisfiability problem is defined as follows: We take $N$
logical variables $x_i\in \{0,1\}$. Then we construct a formula
$F$ of $M$ clauses joined to each other by logical AND. Each
clause contains a number of randomly selected logical variables.
In the random K-SAT problem each clause has a fixed number of $K$
variables. These variables, which join to each other by logical
OR, are negated with probability $1/2$, otherwise appear as such.
For example $F:=( \overline{x}_2\vee x_4)\wedge(\overline{x}_3\vee
x_2)\wedge(x_1 \vee x_3)$ is a 2-SAT formula with $3$ clauses and
$4$ logical variables. A solution of $F$ is a configuration of
logical variables that satisfy all the clauses. The problem is
satisfiable if there is at least one solution or satisfying
configuration of variables for the formula. Given an instance of
the problem, then we are interested to know if it is satisfiable
or not. A more difficult problem is to find the ground
state configurations which lead to the minimum number of violated clauses.\\
The relevant parameter that determines the satisfiability of $F$
is $\alpha:=M/N$. In the thermodynamic limit ($N,M \rightarrow
\infty$ and $\alpha \rightarrow const.$) $F$ is satisfied with
probability one as long as $\alpha < \alpha_c$. Moreover, it has
been found that for $\alpha_d<\alpha<\alpha_c$ the problem is in
the Hard-SAT phase \cite{mez}. At $\alpha_d$ we have a dynamical
phase transition associated with the break down of replica
symmetry. Assuming one-step replica symmetry breaking, one obtains
$\alpha_d \simeq 3.92$ and $\alpha_c\simeq4.26$ for random 3-SAT
problems\cite{mez}. Although this approximation
seems to be exact near the SAT-UNSAT transition but it fails close
to the dynamical transition where higher order replica symmetry
breaking solutions are to be used \cite{mtb,smz}.\\
A useful tool in the study of CSP's is the factor graph which is a
bipartite graph of variable nodes and function nodes (clauses).
The structure of this graph is completely determined by a $M\times
N$ matrix with elements $J_{a,i}\in \{0,+1,-1\}$; $J_{a,i}=+1$ if
clause $a$ contains $x_i$, it is equal to $-1$ if $\overline{x}_i$
appears in $a$ and otherwise $J_{a,i}=0$. In a graph
representation, we add an edges between function node $a$ and
variable node $i$ if $J_{a,i} \ne 0$. The edges will be shown by a
filled line if $J_{a,i}=+1$ and
by a dashed line if $J_{a,i}=-1$.\\
We also define an energy (or cost function) for the problem which
is the number of violated clauses for a given
configuration of variables
\begin{equation}\label{e}
E[\{s\}]\equiv \sum_{a=1}^M \prod_{j=1}^K
\left(\frac{1-J_{a,i_j^a}s_{i_j^a}}{2}\right).
\end{equation}
Here we introduced spin variables $s_i=2x_i-1 \in \{-1,1\}$ and
$i_j^a$ is the index of $j$th variable in clause $a$. A solution
of the problem is a configuration of zero energy and the ground
states are those configuration having the minimum energy. Note
that the presence of two variables in the same clause results to direct
interactions between the corresponding spin variables.

\section{Belief and survey propagation algorithms}\label{3}
In this section we give a brief description of some massage passing
algorithms which help us to get some insights about the solution
space of the problem. These algorithm have an iterative nature and can
give information for single instances of the problem. For more details
about the algorithms and their origin see \cite{kfl,mez,bmz}.\\

\subsection{Assuming replica symmetry; Belief propagation }\label{31}
\begin{figure}
\includegraphics[width=8cm]{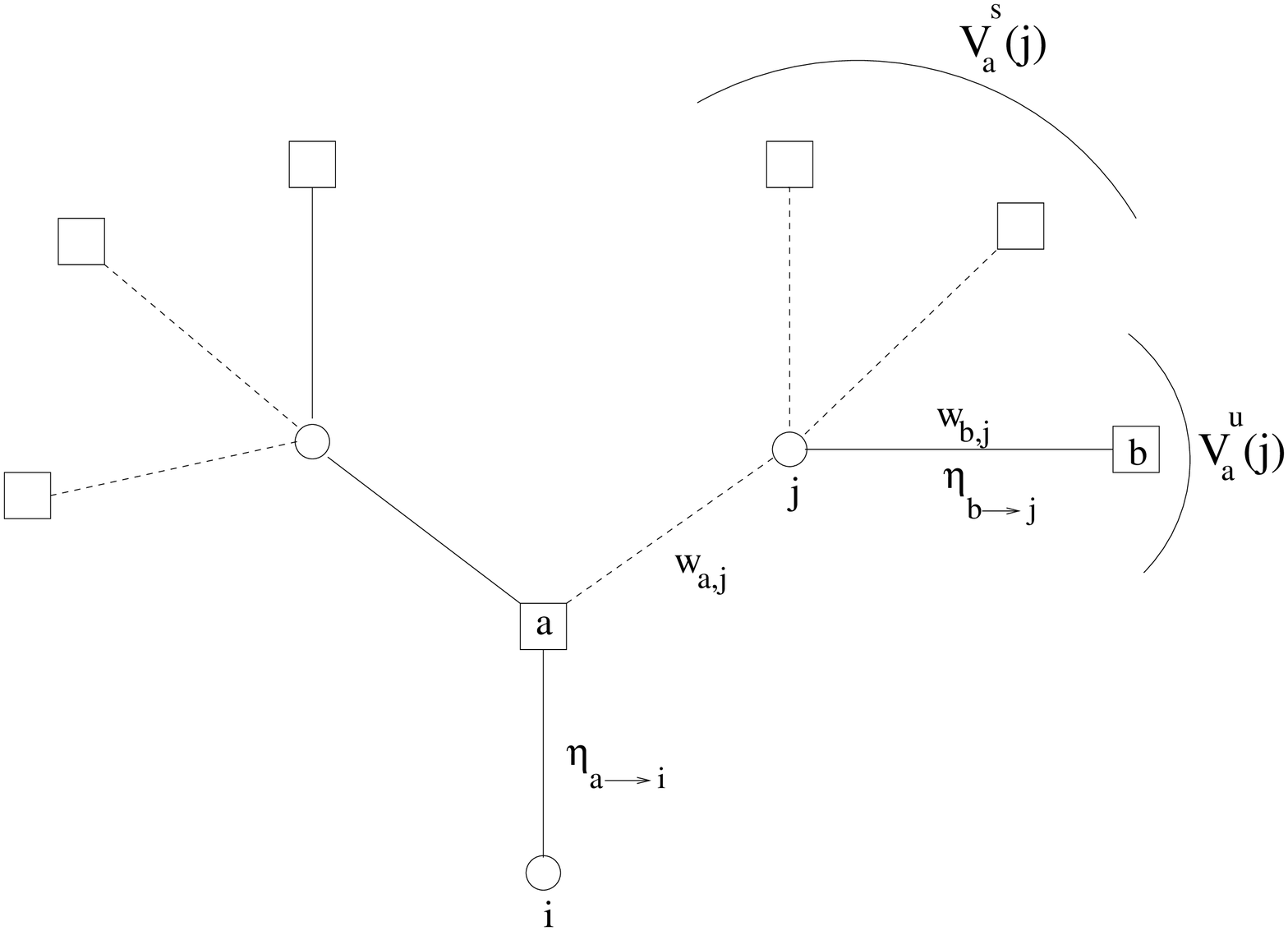}
\caption{Factor graph representation of the problem. Squares and
circles denote to function and variable nodes respectively.
}\label{f1}
\end{figure}

In the following we restrict ourselves in the SAT phase where
there are some solutions that satisfy the problem. These solutions
are represented by points in the $N$-dimensional configuration
space of the variables. If the number of interactions is low enough we
can assume a replica symmetric structure for the organization of
the solutions in this space. It means that the set of solutions
make a single cluster (or pure state) in which any two solutions
can be connected to each other by a path of finite steps when $N$
approaches to infinity. Belief propagation algorithm enables us to
find the solutions and their number (the cluster's size or entropy
of the pure state) in this case. Consider the set of solutions
with $\mathcal{N}_s$ members. Each member is defined by $N$ values for the
variables $\{s_i^*\in \{-1,1\}|i=1,\ldots,N\}$. We consider the
probability space made by all the solutions with equal
probability. Then let us define the warnings $\eta_{a\rightarrow
i}$ as the probability that all variables in clause $a$, except
$i$, are in a state that violate $a$. Assuming a tree-like
structure for the factor graph (i.e. ignoring the correlations
between neighboring variables), $\eta_{a\rightarrow i}$ can be
written as
\begin{equation}\label{eta0}
\eta_{a\rightarrow i}=\prod_{j\in
V(a)-i}P_a^u(j),
\end{equation}
where $P_a^u(j)$ is the probability that variable $j$ dose not
satisfy clause $a$. We also denote by $V(a)$ the set of variables
belong to clause $a$ and by $V(i)$ the set of clauses that
variable $i$ contributes in. In belief propagation algorithm
$P_a^u(j)$ is given by \cite{bmz}
\begin{equation}\label{Pauj}
P_a^u(j)=\frac{\Pi_{j\rightarrow a}^u}{\Pi_{j\rightarrow
a}^s+\Pi_{j\rightarrow a}^u},
\end{equation}
where
\begin{eqnarray}\label{Pii}
\Pi_{j\rightarrow a}^u=\prod_{b\in V_a^s(j)}\left(1-\eta_{b\rightarrow j}\right),\\
\nonumber \Pi_{j\rightarrow a}^s=\prod_{b\in
V_a^u(j)}\left(1-\eta_{b\rightarrow j}\right).
\end{eqnarray}
Here $V_a^s(j)$ denotes to the set of clauses in $V(j)-a$ that
variable $j$ appears in them as it appears in clause $a$, see Fig.
\ref{f1}.
The remaining set of clauses are denoted by $V_a^u(j)$. Starting
from initial random values for $\eta$'s, one can update them
iteratively according to Eqs. \ref{eta0}, \ref{Pauj} and
\ref{Pii}. If the factor graph is spars enough and the problem is satisfiable
then the iteration may converge with no contradictory warnings.
Utilizing these warnings one can use the following relations to
find the entropy of the pure state \cite{bmz}
\begin{equation}\label{comp}
S=\ln \mathcal{N}_s=\sum_{a=1}^M S_a-\sum_{i=1}^N(k_i-1) S_i,
\end{equation}
where
\begin{eqnarray}\label{compai}
S_a=\log[\prod_{j\in V(a)}\left(\Pi_{j\rightarrow
a}^s+
\Pi_{j\rightarrow a}^u\right)-\prod_{j\in V(a)}\Pi_{j\rightarrow a}^u],\\
\nonumber S_i=\log[\Pi_i^-+\Pi_i^+],
\end{eqnarray}
and
\begin{eqnarray}\label{Pi}
\Pi_{i}^-=\prod_{a\in V_+(i)}(1-\eta_{a\rightarrow i}),\\ \nonumber
\Pi_{i}^+=\prod_{a\in V_-(i)}(1-\eta_{a\rightarrow i}).
\end{eqnarray}
In these equations $V_{\pm}(i)$ are the set of function nodes in
$V(i)$
with $J_{a,i}= \pm 1$ and $k_i$ is the number of clauses in $V(i)$.\\
It has been shown that the above algorithm gives exact results for
tree-like factor graphs \cite{bmz}.

\subsection{Assuming one-step replica symmetry breaking; Survey propagation}\label{32}
When we have one-step replica symmetry breaking, the set of
solutions organize in a number of well separated clusters with
their own internal entropies. Suppose there are $\mathcal{N}_c$ of such
clusters. In a coarse grained picture, we can assign a state
$\{\sigma_i^* \in \{-1,0,1\}|i=1,\ldots,N\}$ to each cluster of the
solution space. For a given cluster $\sigma_i^*=+1/-1$ if variable
$i$ has the same value $+1/-1$ in all the solutions belong to the
cluster. Otherwise, that is if variable $i$ is not frozen and
alternates between $-1$ and $1$, $\sigma_i^*=0$. Again we can define
a probability space in which all the clusters have the same
probability. As before $\eta_{a\rightarrow i}$ is the probability
(in new space) that all variables in clause $a$, except $i$, are
in states that violate clause $a$. Notice that we have to take
into account the extra state $\sigma_i^*=0$, which is called the
joker state, in the calculations. Generalizing the belief
propagation relations one obtains \cite{bmz}
\begin{equation}\label{eta1}
\eta_{a\rightarrow i}=\prod_{j\in
V(a)-i}P_a^u(j),
\end{equation}
where
\begin{equation}\label{Pauj1}
P_a^u(j)=\frac{\Pi_{j\rightarrow a}^u}{\Pi_{j\rightarrow
a}^s+\Pi_{j\rightarrow a}^0+\Pi_{j\rightarrow a}^u}.
\end{equation}
But now
\begin{eqnarray}\label{Pii1}
\Pi_{j\rightarrow a}^0=\prod_{b\in
V(j)-a}\left(1-\eta_{b\rightarrow j}\right), \\
\nonumber \Pi_{j\rightarrow a}^u=[1-\prod_{b\in
V_a^u(j)}\left(1-\eta_{b\rightarrow j}\right)]
\prod_{b\in V_a^s(j)}\left(1-\eta_{b\rightarrow j}\right),\\
\nonumber \Pi_{j\rightarrow a}^s=[1-\prod_{b\in
V_a^s(j)}\left(1-\eta_{b\rightarrow j}\right)]\prod_{b\in
V_a^u(j)}\left(1-\eta_{b\rightarrow j}\right).
\end{eqnarray}
The above equations can be solved iteratively for $\eta$'s. As
long as we are in the SAT phase, the above algorithm may
converge with no contradictory warnings. Then
the configurational entropy or complexity of the problem reads
\cite{bmz}
\begin{equation}\label{comp1}
\Sigma=\ln \mathcal{N}_c=\sum_{a=1}^M\Sigma_a-\sum_{i=1}^N(k_i-1)\Sigma_i,
\end{equation}
where
\begin{eqnarray}\label{compai1}
\Sigma_a=\log[\prod_{j\in V(a)}\left(\Pi_{j\rightarrow
a}^s+\Pi_{j\rightarrow a}^0+
\Pi_{j\rightarrow a}^u\right)-\prod_{j\in V(a)}\Pi_{j\rightarrow a}^u],\\
\nonumber \Sigma_i=\log[\Pi_i^-+\Pi_i^0+\Pi_i^+],
\end{eqnarray}
and
\begin{eqnarray}\label{Pi1}
\Pi_{i}^-=[1-\prod_{a\in V_-(i)}(1-\eta_{a\rightarrow
i})]\prod_{a\in V_+(i)}(1-\eta_{a\rightarrow i}),\\ \nonumber
\Pi_{i}^+=[1-\prod_{a\in
V_+(i)}(1-\eta_{a\rightarrow i})]\prod_{a\in V_-(i)}(1-\eta_{a\rightarrow i}),\\
\nonumber \Pi_{i}^0=\prod_{a\in V(i)}(1-\eta_{a\rightarrow i}).
\end{eqnarray}
To find a solution of the problem we can follow a simple survey
inspired decimation algorithm that works with the biases a
variable experience \cite{bmz}. Let us define $W_i^+$ the
probability for variable $i$ to be frozen in state $+1$ in a
randomly selected cluster of solutions. Similarly we define
$W_i^-$ and $W_i^0$. Then according to the above definitions we
have
\begin{eqnarray}\label{Wi}
W_i^+=\frac{\Pi_{i}^+}{\Pi_{i}^++\Pi_{i}^0+\Pi_{i}^-},\\ \nonumber
W_i^-=\frac{\Pi_{i}^-}{\Pi_{i}^++\Pi_{i}^0+\Pi_{i}^-},\\ \nonumber
W_i^0=1-W_i^+-W_i^-.
\end{eqnarray}
After a run of survey propagation algorithm we have the above
biases and fix the most biased variable, i.e. one with largest
$|W^+-W^-|$. Then we can simplify the problem and again run survey
propagation algorithm. We repeat the above process until we reach
a simple problem with all warnings equal to zero. This problem
then can be solved by a local search algorithm.

\section{Finding satisfiable subproblems}\label{4}
Consider a satisfiable random KSAT problem and the associated
factor graph with $N$ variable nodes, $M$ function nodes and $KM$
edges. All the function nodes have the same degree $k_a=K$ and a
variable node has degree $k_i$ which, in the thermodynamic limit,
follows a Poisson distribution of mean $K\alpha$. If
$\{s_i^*|i=1,\ldots,N\}$ is a solution of the problem, then any
function node in the factor graph has at least one neighboring
variable node that satisfies it. It means that for any solution we
can remove some of the edges in the factor graph while the
obtained subproblem is still satisfiable and  $k_a\ge 1$ for all
function nodes. Obviously we can do this until each function node
is only connected to  one variable node, the one that satisfies
the corresponding clause. So it is clear that for a satisfiable
problem there exist many subproblems ranging from the original
one, with $L=KM$ edges (or interactions), to a minimal one with
$L=M$ edges in its factor graph. In general we define $G_x(M,N)$
as the ensemble of satisfiable subproblems defined by the
parameter $x$. For example $G_L(M,N)$ is the ensemble of
satisfiable subproblems with $L$ edges. \\
An interesting point is the presence of a correspondence between
the solutions of the original problem and solutions of an ensemble
of subproblems with $L$ edges. Obviously any solution of the
subproblems in $G_L(M,N)$ is also a solution of the original
problem. Moreover, as described above, for any solution we can
remove some of the edges until we obtain a subproblem of exactly
$L$ edges. In \cite{rm} we showed that this correspondence holds
also for the set of spanning trees which is a subset of
$G_{M+N-1}(M,N)$.  These correspondence relations will allow us to
construct the ensembles and to find the solutions of the original
problem by solving a subproblem.\\
Notice that as the number of interactions in a problem decreases we have to pay less
computational cost to solve it. In fact, tree-like factor
graphs can easily be solved by efficient local search algorithms.
And if someone could give the ensemble of
minimal subproblems, the whole set of solutions would be available.
Now the main questions are:  How can we construct these
satisfiable subproblems and what can be said about the properties
of these subproblems? In the following we try to answer these
questions by a simple modification of survey
propagation algorithm, introduced in \cite{rm}.\\

\subsection{General arguments}\label{41}
For a given ensemble of subproblems $G_x(M,N)$ we would have
$\mathcal{N}_x(M,N)$ members.
In a given ensemble $G_x$, we assign weight $w_{a,i}$ to edge
$(a,i)$ as a measure of its appearance frequency in the ensemble,
that is,
\begin{equation}\label{wai}
w_{a,i}=\frac{1}{\mathcal{N}_x}\sum_{g \in G_x} y_{a,i}(g),
\end{equation}
where $y_{a,i}(g)=1$ if the edge appears in $g$ and otherwise
$y_{a,i}(g)=0$. Let $P_x(g)$ be a measure defined on the space of
all subgraphs with equal probability for all subgraphs $g$ that
belong to $G_x$ and zero otherwise. This probability can be
written in terms of $y$'s
\begin{equation}\label{PLg}
P_x(g)=\frac{1}{\mathcal{N}_x}\sum_{g' \in G_x}
\prod_{(a,i)}\delta_{y_{a,i}(g),y_{a,i}(g')}.
\end{equation}
It is then easy to show that $\sum_g P_L(g)=1$ and $\ln
\mathcal{N}_x=-\sum_g P_x(g) \ln P_x(g)$. Suppose that we have
obtained $w$'s for the ensemble $G_x$ from another way. As an
estimate of $P_x(g)$ we write
\begin{equation}\label{PLge}
P_x^{e}(g)= \prod_{(a,i)}
[y_{a,i}(g)w_{a,i}+(1-y_{a,i}(g))(1-w_{a,i})].
\end{equation}
Then we expect that
\begin{equation}\label{NLe}
\ln \mathcal{N}_x^e=-\sum_g P_x^e(g) \ln P_x^e(g)=- \sum_{(a,i)}
[w_{a,i}\ln w_{a,i}+(1-w_{a,i})\ln(1-w_{a,i})],
\end{equation}
gives a good estimate of $\ln \mathcal{N}_x$.\\
Suppose that we have obtained all the members in ensemble $G_x$.
Assuming replica symmetry, we could run belief propagation on each
member of the ensemble and obtain its entropy. Then we could
define $<S_x>$, the average of entropy taken over ensemble $G_x$.
Similarly we could run survey propagation algorithm and define
$<\Sigma_x>$ as the average complexity of subproblems in $G_x$.
Actually we will not follow the above procedure and get around the
difficult problem of finding all the ensemble members. Let us
describe our procedure for the case of survey propagation
algorithm. Generalization to the belief propagation
algorithm would be straightforward.\\
To obtain $w$'s for an ensemble we go through a self-consistency
approach. We run survey propagation algorithm on the original
factor graph but at the same time we take into account the fact
that each edge has its own probability of appearing in the
ensemble. Now the survey along edge $(a,i)$ is updated according
to the following rule
\begin{equation}\label{eta2}
\eta_{a\rightarrow i}=\prod_{j\in
V(a)-i}[w_{a,j}P_a^u(j)+1-w_{a,j}],
\end{equation}
where as before $P_a^u(j)$ is given by Eq.\ref{Pauj1} with
\begin{eqnarray}\label{Pii2}
\Pi_{j\rightarrow a}^0=\prod_{b\in
V(j)-a}\left(1-w_{b,j}\eta_{b\rightarrow j}\right), \\
\nonumber \Pi_{j\rightarrow a}^u=[1-\prod_{b\in
V_a^u(j)}\left(1-w_{b,j}\eta_{b\rightarrow j}\right)]
\prod_{b\in V_a^s(j)}\left(1-w_{b,j}\eta_{b\rightarrow j}\right),\\
\nonumber \Pi_{j\rightarrow a}^s=[1-\prod_{b\in
V_a^s(j)}\left(1-w_{b,j}\eta_{b\rightarrow j}\right)]\prod_{b\in
V_a^u(j)}\left(1-w_{b,j}\eta_{b\rightarrow j}\right).
\end{eqnarray}
An essential step here is the determination of $w$'s in a given
ensemble. Remember that a given ensemble is a set of satisfiable
subproblems which completely define the probabilities $w$ along
the edges of the factor graph. Thus, if with a given set of
$w$'s we find a large warning sent from $a$ to $i$,  we expect a
high probability for the presence of that edge in the ensemble.
Here we make a crucial assumption and use the following ansatz
\begin{equation}\label{weta}
w_{a,i}=[\eta_{a\rightarrow i}]^{\mu}.
\end{equation}
that incorporates the above fact. We take $\mu \ge 0$ as a free
parameter and denote the resulted ensembles by $G_{\mu}$.
For a given $\mu$ we would have an ensemble of satisfiable
subproblems with different number of edges. Because of the
functional form of the above ansatz, the average number of edges
in the ensemble decreases by increasing $\mu$. Therefore, to
obtain smaller satisfiable subproblems we will need to run the
algorithm for larger values of $\mu$.\\
Starting from initially random $\eta$'s and $w$'s we iterate the
above equations until (i) it converges to some fixed point, (ii)
or results to contradictory warnings (iii) or dose not converge
in a predefined limit for the number of iterations $t_{max}$. We
think that as long as the original problem is satisfiable the
algorithm will converge in a finite fraction of times that we run
it.\\
If the algorithm converges then we can utilize our definition for
$w$'s and construct satisfiable subproblems. To construct a
subproblem in $G_{\mu}$ we go through all the edges and select them with probabilities  $w_{a,i}$'s. We hope that such a
subproblem be satisfiable with a considerable probability.
Moreover, it is reasonable that we pay more computational cost to
find smaller satisfiable subproblems which are closer to the
solutions of the original problem.

\subsection{Numerical results}\label{42}
In the following we will study some properties of satisfiable subproblems including the spanning trees of
the original factor graph and the minimal subproblems.\\
We start from initially random values of $0\le
\eta_{a,i},w_{a,i}\le 1 $ for all the edges $(a,i)$. Then in each
iteration of the algorithm we update $\eta$ and $w$ for all the
edges according to Eqs. \ref{eta2}, \ref{Pii2} and \ref{weta}. The
edges are selected sequentially in a random way. The algorithm
converges if for all the edges the differences between new and old
values of $\eta$ are less than $\epsilon$. We bound the number of
iterations from above to $t_{max}$ and if the algorithm dose not
converge in this limit, we say that it diverges. In the following
we will work with $\epsilon=0.001$
and $t_{max}=1000$. Moreover, we consider $3$-SAT problems where
each clause in the original problem has just $K=3$ variables.\\
\begin{figure}
\includegraphics[width=8cm]{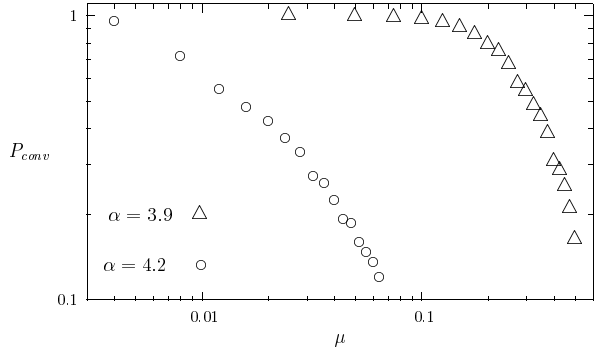}
\caption{Convergence probability for two values of $\alpha$ close
to the SAT-UNSAT transition. Number of variables is $N=1000$ and
statistical errors are of order $0.01$.}\label{f2}
\end{figure}
\begin{figure}
\includegraphics[width=8cm]{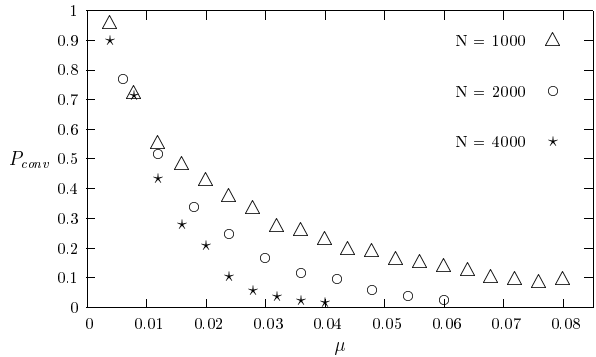}
\caption{Convergence probability for different problem sizes at
$\alpha=4.2$. Statistical errors are of order $0.01$.}\label{f3}
\end{figure}
Let us first study the convergence properties of the modified
algorithm. To this end we repeat the algorithm for a number of
times and define $P_{conv}$ as the fraction of times in which the
algorithm converges. In Fig. \ref{f2} we display $P_{conv}$ for
the modified survey propagation algorithm. It is observed that
$P_{conv}$ decreases by increasing $\mu$. Moreover, $P_{conv}$
diminishes more rapidly for larger $\alpha$. It is reasonable
because the removal of edges becomes harder as we get closer to
the SAT-UNSAT transition. What happens if we increase the problem
size? Figure \ref{f3} shows the finite size effects on convergence probability.
These effects are significant due to the small problem sizes studied here.
Moreover, as expected, the probability decreases more rapidly as $N$ increases.\\
\begin{figure}
\includegraphics[width=8cm]{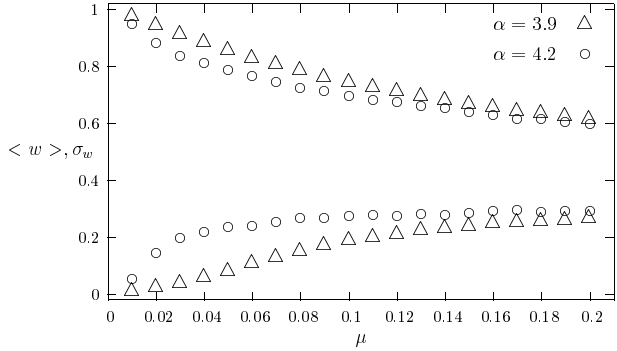}
\caption{The average weight and its standard deviation for
$N=1000$. Statistical errors are about the point's sizes.
}\label{f4}
\end{figure}
To see how the number of edges changes with $\mu$ we obtained the
average weight of an edge, $<w>$, and its standard deviation,
$\sigma_w$, in converged cases. The average number of edges is
given by $<L>=3M<w>$. Fig. \ref{f4} shows how these quantities
behave with $\mu$. We found that as $\mu$ gets larger $<w>$
decreases and finally (not shown in the figure) approaches to
$1/3$, the minimum possible value to have a satisfiable subproblem
when $K=3$.\\
\begin{figure}
\includegraphics[width=8cm]{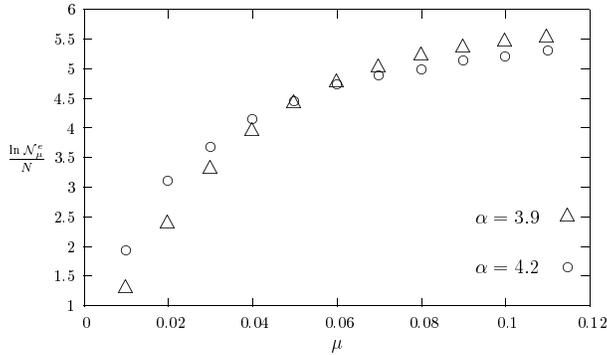}
\caption{Estimated value of the number of members in $G_{\mu}$.
The results are for $N=1000$ and statistical errors are of order
$0.1$. }\label{f5}
\end{figure}
Using our arguments in previous subsection we can obtain an
estimate of the number of members in the
ensemble $G_{\mu}$, $\mathcal{N}_{\mu}^e$. In Fig. \ref{f5} we show how $\ln
\mathcal{N}_{\mu}^e$ changes with $\mu$. Here we have displayed
the results just for small $\mu$'s where we are interested in. For
larger $\mu$'s, $\mathcal{N}_{\mu}^e$
decreases to its value for $<L>=M$.\\
\begin{figure}
\includegraphics[width=8cm]{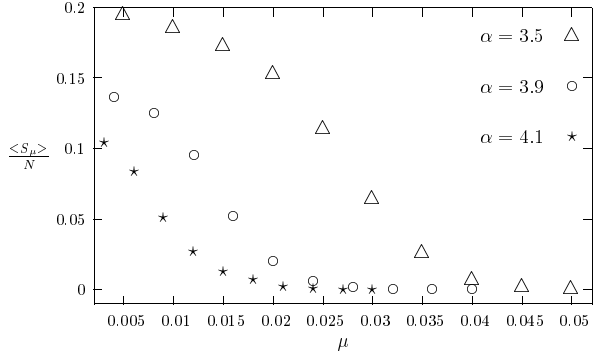}
\caption{Average entropy of a subproblem in $G_{\mu}$ for
$N=1000$. Statistical errors are about the point's
sizes.}\label{f6}
\end{figure}
\begin{figure}
\includegraphics[width=8cm]{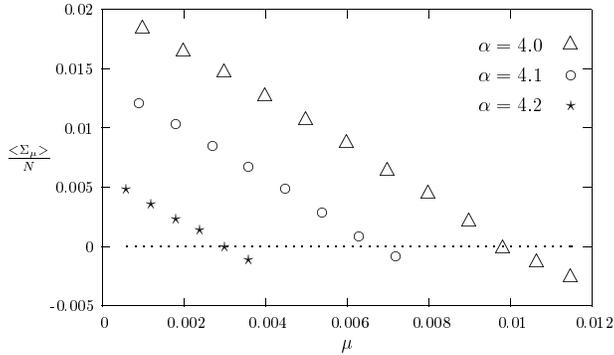}
\caption{Average complexity of a subproblem in $G_{\mu}$ for
$N=1000$. Statistical errors are about the point's
sizes.}\label{f7}
\end{figure}
As described in the previous section we can obtain the average
entropy of a typical subproblem in $G_{\mu}$ by running belief
propagation on it. The results have been displayed in Fig.
\ref{f6}. Similarly the average complexity of a subproblem is
obtained by running survey propagation algorithm. Figure \ref{f7}
shows this quantity for some values of $\alpha$. As the figures
shows both $<S_{\mu}>$ and $<\Sigma_{\mu}>$ diminish with $\mu$
and $\alpha$; Removing edges from the factor graph and approaching
the SAT-UNSAT transition both decrease the number of solutions and
complexity. Notice that for a fixed value of $\alpha$ we can
define the threshold $\mu_c(\alpha)$ where the complexity
vanishes.
It is a decreasing function of $\alpha$ and we know already that $\mu_c(\alpha_c)=0$.\\

\subsection{Satisfiable spanning trees}\label{43}
\begin{figure}
\includegraphics[width=8cm]{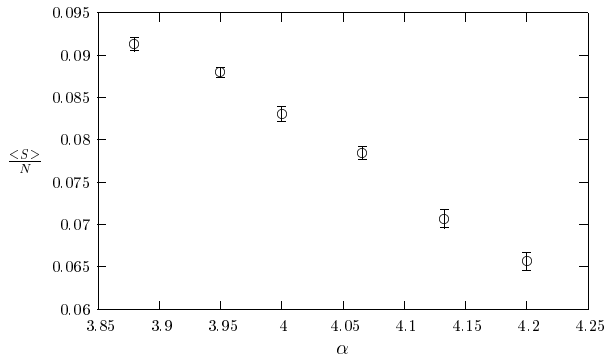}
\caption{Entropy of typical satisfiable spanning trees for $N=1000$ and $\mu=0.04$. }\label{f10}
\end{figure}

\begin{figure}
\includegraphics[width=8cm]{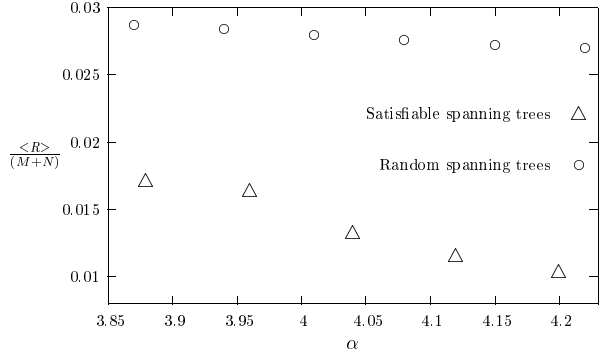}
\caption{Average diameter of satisfiable and random spanning trees for  $N=1000$
and $\mu=0.05$. }\label{f12}
\end{figure}

Suppose that the algorithm converges and returns the weights $w$'s
for all the edges of the factor graph. It is not difficult to
guess that maximum spanning trees have a larger probability to be
a satisfiable spanning tree. A maximum spanning tree is a spanning
tree of the factor graph with maximum weight
$W=\sum_{(a,i)}w_{a,i}$.  For a given $\mu$ and a converged case
we can construct maximum spanning trees in the following way: We
start from a randomly selected node in the original factor graph
and find the maximum weight among the edges that connect it to the
other nodes. Then we list the edges having a weight in the
$\epsilon$-neighborhood of the maximum one and add randomly one of
them to the new factor graph. If we repeat the addition of edges
$N+M-1$ times we obtain a spanning tree factor graph which has the
maximum weight on its edges. Notice that taking a nonzero interval
to define the edges of maximum weight at each step, along with the
randomness in choosing one of them, allow to construct a large
number of maximum spanning trees. In this way we define $P_{sat}$
as the probability that a maximum spanning tree be satisfiable if
the algorithm converges. To find out the satisfiability of the
subproblem we use a local search algorithm (Focused Metropolis
Search) introduced in \cite{sao}.
\begin{figure}
\includegraphics[width=8cm]{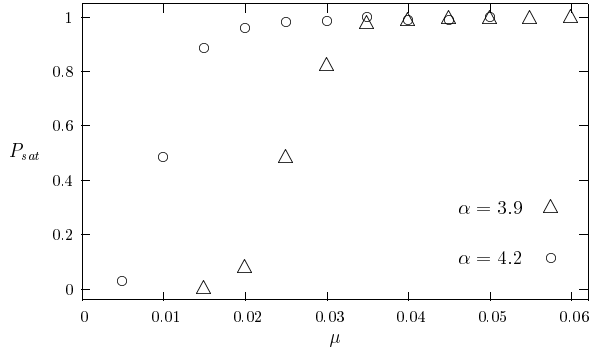}
\caption{Satisfiability probability of maximum spanning trees
versus $\mu$. The problem size is $N=1000$ and statistical errors
are of order $0.01$.}\label{f8}
\end{figure}
Figure \ref{f8} displays this quantity versus $\mu$ for some
values of $\alpha$. The probability to find a satisfiable spanning
tree becomes considerable even for a very small $\mu$ and finally
approaches to $1$. For instance, if $\alpha=4.2$ then at $\mu=0.01$
almost half the maximum spanning trees are satisfiable. For these
parameters the fraction of converged cases is nearly $0.6$ (see
Fig. \ref{f2}). Although the algorithm provides a simple way of
constructing satisfiable spanning trees but in general finding
them is not an easy task. For example for a satisfiable problem
with parameters $(N=100,M=400,K=3)$, we found no satisfiable
spanning tree among $10^7$ randomly constructed ones.
\begin{figure}
\includegraphics[width=8cm]{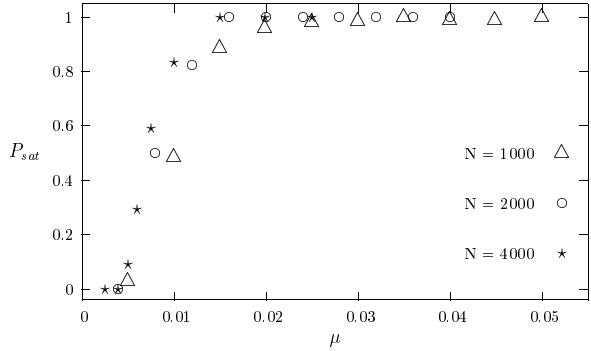}
\caption{Satisfiability probability of maximum spanning trees for
a few problem sizes. Statistical errors are of order
$0.01$.}\label{f9}
\end{figure}
Figure \ref{f9} shows the satisfiability of maximum spanning trees
for some larger problem sizes at $\alpha=4.2$. Hopefully, by
increasing $N$ the satisfiability probability enhances for smaller
values of $\mu$ and gets more rapidly its saturation value. We
hope that this behavior of $P_{sat}$ compensate the decrease in
$P_{conv}$ for larger problem sizes. A look at Figs. \ref{f3} and
\ref{f9} shows that for $N=4000$, $\alpha=4.2$ and at $\mu=0.01$
we have $P_{conv} \approx 0.5$ and $P_{sat}\approx 0.7$. It means
that of $100$ runs we can extract on average $35$ satisfiable
spanning trees. Having a satisfiable spanning tree then we can
find its solutions (which are also the solutions of the original
problem) by any local search algorithm. This, besides the other
methods, provides another way of finding the solutions of the
original problem. In Fig. \ref{f10} we obtained the entropy of
typical satisfiable spanning trees by running belief propagation
on them. As the figure shows this entropy decreases linearly with
$\alpha$.
\begin{figure}
\includegraphics[width=8cm]{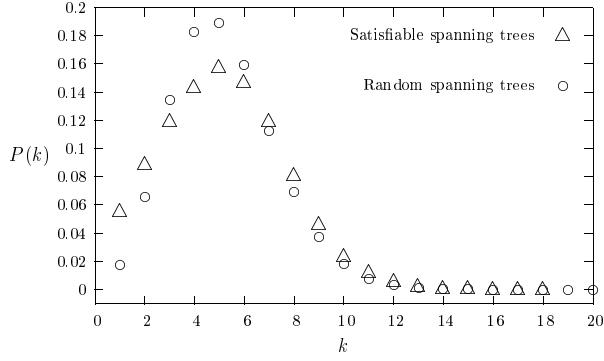}
\caption{Degree distribution of variable nodes in satisfiable and
random spanning trees. The parameters are $N=1000$, $\alpha=4.2$
and $\mu=0.025$. Statistical errors are about the point
sizes.}\label{f11}
\end{figure}
It will be interesting to compare the structural properties of
satisfiable spanning trees with those of randomly constructed
ones. To this end we obtained the degree distribution of variable
and function nodes in the corresponding spanning trees. In Fig.
\ref{f11} we compare the degree distributions of variables. For
function nodes we found no significant difference between the two
kinds of spanning trees. However, the degree distribution of
variable nodes is slightly broader for the satisfiable spanning
trees. There are more low and high degree nodes in these spanning
trees. Another feature of satisfiable spanning trees is their low
diameter compared to the random ones; Take the node having
maximum degree as the center of spanning tree. The distance of a
node from the center is defined as the number of links in the
shortest path connecting the center to the node. We define the
largest distance in the network as its diameter. The diameter of
the two sets of spanning trees has been compared in Fig.
\ref{f12}. Satisfiable spanning trees have a diameter which is
almost half the diameter of the random spanning trees.\\

\subsection{Minimal satisfiable subproblems}\label{44}
\begin{figure}
\includegraphics[width=8cm]{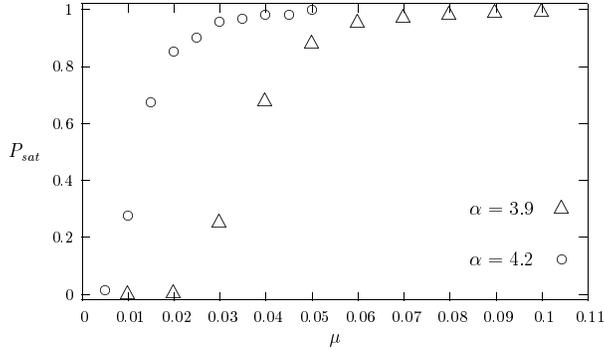}
\caption{Satisfiability probability of a minimal subproblem versus $\mu$. Number of
variables is $N=1000$ and statistical errors are of order $0.01$.}\label{f13}
\end{figure}

\begin{figure}
\includegraphics[width=8cm]{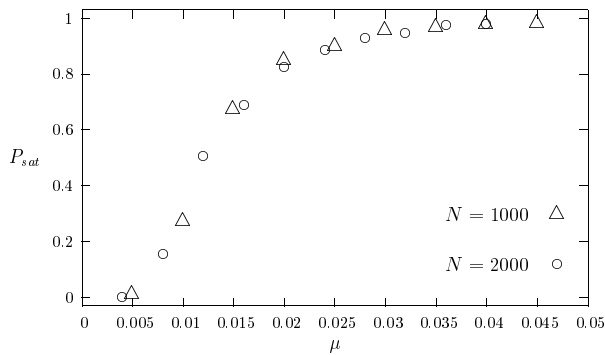}
\caption{Satisfiability probability of the minimal subproblems at $\alpha=4.2$.
Statistical errors are of order $0.01$.}\label{f14}
\end{figure}

\begin{figure}
\includegraphics[width=8cm]{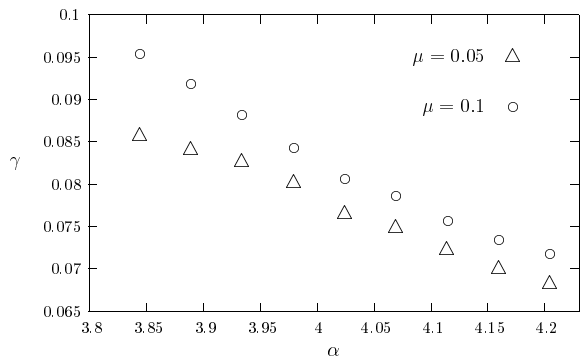}
\caption{Fraction of free variables in the minimal satisfiable
subproblems. Number of variables is $N=1000$ and statistical
errors are of order $0.001$.}\label{f15}
\end{figure}

\begin{figure}
\includegraphics[width=8cm]{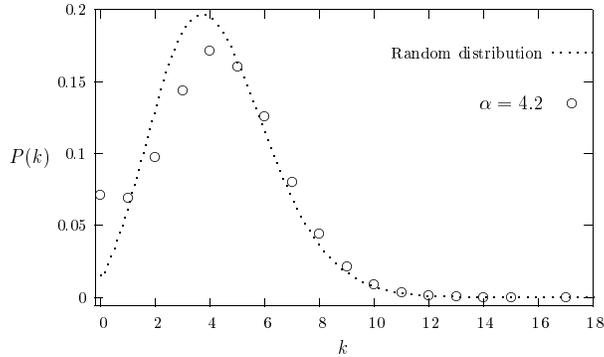}
\caption{Comparing degree distribution of variable nodes in
satisfiable and random minimal subproblems. Number of variable
nodes is $N=1000$ and statistical errors are of order
$0.01$.}\label{f16}
\end{figure}

A minimal subproblem has the minimum possible number of edges $L=M$
where each function node is connected to at most one variable
node. Having such a subproblem it is easy to check its
satisfiability. The solutions of a minimal satisfiable subproblem
will be the solutions of the original problem. Moreover for any
solution of the original problem there is at least one minimal
satisfiable subproblem. The total number of minimal subproblems is
$K^M$ that makes the exhaustive search among them for satisfiable
ones an intractable task.\\
Suppose that the algorithm for a given $\mu$ has been converged
and returned the weights $w$'s for all the edges. Among the edges
emanating from function node $a$ we choose the one with maximum
weight. If there are more than one edge of maximum weight then we
select one of them randomly. Notice that we treat all the edges in
the $\epsilon$-neighborhood of the maximum weight in the same
manner. For all the function nodes we do the above choice to
construct a minimal subproblem. Then we check the satisfiability
of the subproblem and repeat the process for a large number of
minimal subproblems obtained from converged runs of the algorithm.
We define $P_{sat}$ as the probability that a minimal subproblem
be a satisfiable one. This quantity has been displayed in Fig.
\ref{f13}.
Again we see that even for very small $\mu$, $P_{sat}$ is close to
$1$. When the parameters are $(N=1000,M=4200,K=3)$ this happens at
$\mu \approx 0.05$. According to Fig. \ref{f2}, at these
parameters we have to run the algorithm on average $15$ times
to find a converged case. In Fig. \ref{f14} we compare $P_{sat}$ for
two different problem sizes. As the figure shows there is no significant
difference between the two results.\\
Having a minimal satisfiable subproblem we will be able to find
the solutions directly. Any variable node that has at least one
emanating edge is frozen in the obtained set of solutions. In Fig.
\ref{f15} we have showed the fraction of free variables versus
$\alpha$. Notice that $1-\gamma$ is the fraction of frozen
variables and $2^{N\gamma}$ gives the number of solutions in a
typical satisfiable subproblem. As expected the number of frozen
variables increases as we get closer to the SAT-UNSAT transition.
Finally we look at the degree distribution of variable nodes in
the minimal satisfiable subproblems. In Fig. \ref{f16} we compare
the degree distribution at $\alpha=4.2$ with the random case in
which the edges have been distributed randomly while the other
parameters are the same.  We observe that the real distribution is
broader than the random one. Low and high degree nodes have more
contribution in  the minimal satisfiable subproblems. We
encountered the same phenomenon in Fig. \ref{f11} that compares
degree distribution of satisfiable spanning trees with the random
ones.

\section{Conclusion}\label{5}

In summary we showed that there is a way to reduce a random
K-satisfiability problem to some simpler subproblems their
solutions are also the solutions of the original problem. To
achieve this we modified the known message passing algorithms by
assigning some weights to the edges of the factor graph. Finding
satisfiable subproblems allowed us to compute the expected value
of their entropy and complexity. In the case of satisfiable
spanning trees we could compare their structural properties with
those of random spanning trees. We could also construct the minimal
satisfiable subproblems and study some interesting features of their factor graph.\\
The modified algorithm studied in this paper can be used, besides the the present algorithms, to find the solutions of a constrained
satisfaction problem in the SAT phase. Moreover, it provides a way
to find the satisfiable subproblems which is not an easy
task. Comparing satisfiable subproblems with equivalent random
ones might provide some insights about the nature of satisfiable
problems and so their solutions.\\
Due to the computational limitations, the results have been restricted
to small problem sizes of order $10^3$. In this paper we tried to show the trend by studding
different problem sizes.

\end{document}